\documentclass[a4paper]{isp}
\usepackage{graphicx,url}
\def\aj{AJ  }
\def\apj{ApJ\,  }
\def\apjl{ApJ\,  }
\def\apjs{ApJS  }
\def\apss{Astrophysics and Space Science  }

\def\mnras{MNRAS\,  }

\def\aj{AJ  }
\def\apj{ApJ\,  }
\def\apjl{ApJ\,  }
\def\apjs{ApJS  }
\def\apss{Astrophysics and Space Science  }

\def\mnras{MNRAS\,  }

\begin{document}
\def\h0units{\mathrm{km\,s^{-1}\,Mpc^{-1}}}
\def\cunits{\mathrm{km\,s^{-1}}}
\def\lunits{\mathrm{erg\,s^{-1}}}
\newcommand{\dl}{D_{L}}
\newcommand{\pl}{D_{L,2,2}}
\newcommand{\ml}{D_{L,m,2,2}}
\newcommand{\om}{\Omega_{\rm M}}
\newcommand{\ok}{\Omega_K}
\newcommand{\ola}{\Omega_{\Lambda}}
\def\sun{\hbox{$\odot$}}
\title
{
A lognormal luminosity function
for SWIRE in  flat cosmology
}

\author{Lorenzo Zaninetti $^{1,}$}

\institute
{
$^{1}$
Physics Department,
 via P.Giuria 1,\\ I-10125 Turin,Italy
}
\maketitle

\begin{abstract}
The evaluation  of the physical parameters
or effects---such as the luminosity function (LF)
or  the photometric maximum (PM)---for galaxies is routinely modeled by their
spectroscopic redshift.
Here, we model LF and PM for galaxies by the
photometric redshift as given by the 
Spitzer Wide-Area Infrared Extragalactic
(SWIRE) catalog
in the framework of a lognormal LF.
In addition, we compare our model with the Schechter LF for galaxies.
The adopted cosmological framework is that of the flat cosmology.
\end{abstract}
\begin{keywords}
{
galaxy groups, clusters, and superclusters; large scale structure of the Universe
Cosmology
}
\end{keywords}

\section{Introduction}

When a spectrum is not available
for direct redshift determination, the photometric redshift
of a galaxy can be deduced  from
its colors.
In practice, the galaxy's   magnitude in several broad-band
filters is compared to that expected from the
theoretical spectra of different types of
galaxies at a range of redshifts.
In the last few years, the number of catalogs characterized by
the  photometric redshift as an indicator of the distance has grown progressively.
For example, the Spitzer Wide-Area Infrared Extragalactic 
survey (SWIRE) catalog, see \cite{Rowan-Robinson2013},
has one million galaxies while the
GLADE catalog, see \cite{Dalya2016}, has
one and a half million galaxies.
The lognormal distribution in astronomy has 
been  used to model 
the apparent distribution
of galaxies,
see \cite{Karasev1982},
the 
durations of gamma-ray
burst (GRB), 
see \cite{Mcbreen1994,Li1996,Nakar2001,Ioka2002},
the  luminosity function (LF) of GRB, 
see \cite{Zaninetti2016c}, 
the time interval between successive bursts from 
the magnetar SGR 1806-20,
see \cite{Hurley1995,McBreen1998},
the angular momentum of disc galaxies and 
the galaxies rotation curve,
see \cite{Marr2015a,Marr2015b}.
The  LF for galaxies
is usually fitted with  the Schechter function,
see  \cite{schechter}.
A first  improvement for  the standard  LF  can be obtained
from a given
probability density function  (PDF) (i.e. the gamma variate)
for the mass of the galaxies, $\mathcal { M }$,
and by
assuming  a non-linear relationship
between ${\mathcal M}$ and  luminosity (L), see \cite{Zaninetti2008}.
A second improvement analyzes standard PDFs, such as
the generalized gamma,  and then converts it to a LF,
see \cite{Zaninetti2010f}.
The  newly obtained LFs for galaxies should then be compared
with the Schechter LF in the framework of the statistical
tools.
The  high number of galaxies     allows us to pose the following
questions:
\begin{enumerate}
\item
 What are the  differences  between the photometric redshift (PM) and the spectroscopic
   redshift~?
\item
 Can we model  the number of galaxies as a function of the
   PM? For example, see Figure 15 in
   \cite{Bilicki2016}~?
\item
 Can we model the  LF  for galaxies with the  lognormal distribution 
 when the photometric redshift is available~?
\end{enumerate}

The rest of this paper is structured as follows. In Section \ref{secflatcosmo}, we
explore
the luminosity distance as a function of the redshift
in flat cosmology.
In Section \ref{sec_lf},
we evaluate  the astronomical LF for galaxies
and we then model it with the lognormal LF.
In Section \ref{sec_maximum},  we model the
maximum number of galaxies (PM) in the SWIRE catalog
as a function of the
photometric redshift.

\section{Flat cosmologies with a cosmological constant}
\label{secflatcosmo}

The luminosity distance $dl$ is
\begin{equation}
  \dl(z;c,H_0,\om) = \frac{c}{H_0} (1+z) \int_{\frac{1}{1+z}}^1
  \frac{da}{\sqrt{\om a + (1-\om) a^4}} \quad ,
  \label{lumdistflat}
\end{equation}
where $H_0$
is the Hubble constant, expressed in     $\h0units$;
$c$ is the light velocity,  expressed in $\cunits$;
$z$ is the redshift;
$a$ is the scale-factor;
and  $\om$ is
\begin{equation}
\om = \frac{8\pi\,G\,\rho_0}{3\,H_0^2}
\quad ,
\end{equation}
where $G$ is the Newtonian gravitational constant and
$\rho_0$ is the mass density at the present time.

We report the Pad\'e approximate integral
of the luminosity distance
$\pl$, in the case of  $m$=2 and $n$=2 when
$H_0 = 70  \h0units$ and $\om=0.277$ as  in
\cite{Varela2012}
\begin{eqnarray}
\pl(z)
=
 4282.749\,   ( 1+z   )    ( - 0.07115\,   ( 1+z
   ) ^{-1}+ 0.12536\,\ln    ( - 699.225\,   ( 1+z
   ) ^{-2}
\nonumber \\
+ 124.1677\,   ( 1+z   ) ^{-1}
- 282.588
   ) + 1.32386- 0.39385\,i
\nonumber \\
- 2.17281\,\arctan   (
 1.58858\,   ( 1+z   ) ^{-1}- 0.141049   )
   )
\quad .
\label{dlpade22}
\end{eqnarray}
In this  complex analytical solution, we have   a real part
that is denoted by $\Re$
and a negligible imaginary part.
For example, this real part is
35089.79318   when $z=4$
and the imaginary  part is $-0.706\,10^{-4}$.

In the case of $m$=2 and $n$=2 the minimax rational expression
for the luminosity distance, $\ml$,
is
\begin{eqnarray}
\ml =
\frac 
{ 
2.982+ 1868.83\,z+ 2950.527\,{z}^{2}
}
{
0.453585+ 0.26391\,z+ 0.0030963\,{z}^{2}
}
\\
for  \quad 0.0001 <z< 4 
\nonumber  
\quad .
\label{dlminimax}
\end{eqnarray}
These formula can be inverted
for the redshift, $z_{2,2}(\dl)$,
\begin{equation}
z_{2,2}(\dl)
=
\frac{N22}{D22}
\label{z22}
\quad ,
\end{equation}
where
\begin{equation}
N22 = 
- 2.686\,10^{10}\,{\it \dl}+ 1.292\,10^{14}- 20 \,\sqrt {
 1.684\,10^{18} {{\it \dl}}^{2}+{ 7.283 \, 10^{22
}}\,{\it \dl}+{ 4.141 \, 10^{25}}}
\quad , 
\end{equation}
and
\begin{equation}
D22 =
5.4075\,10^8\,{\it \dl}- 4.041511\,10^{14}
\quad  .
\end{equation}

The  angular
diameter distance, $D_A$, is a second useful distance; which,
after \cite{Etherington1933},
is
\begin{equation}
D_A = \frac{D_L}{(1+z)^2}
\quad .
\end{equation}
The  transverse comoving distance, $D_M$, is a third  useful distance,
\begin{equation}
D_M = \frac{D_L} {1+z}
\quad ,
\label{comovingdistance}
\end{equation}
with the connected  total comoving volume
$V_c$
\begin{equation}
V_c= \frac{4}{3}\pi D_M^3
\quad .
\label{comovingvolume}
\end{equation}

\section{The luminosity function}
\label{sec_lf}

This section  introduces the SWIRE catalog. It
discusses  the differences between photometric and
spectroscopic redshift,
evaluates the astronomical LF,
introduces the Schechter   LF,
and models the results with the lognormal LF.

\subsection{The SWIRE catalog}

The SWIRE
photometric redshift catalog
contains over 1 million  galaxies
over   49 $deg^2$ of sky.
The parameters that are used here are:
the   absolute B-band magnitude;
the luminosity in the B-band, which is
expressed in solar units; and, the photometric redshift,
see \cite{Rowan-Robinson2013}
with data  at
\url{http://vizier.u-strasbg.fr/viz-bin/VizieR}
and  specifically Table II/326/zcatrev.

\subsection{Photo-z vs. spectro-z}

\begin{figure}
\begin{center}
\includegraphics[width=7cm]{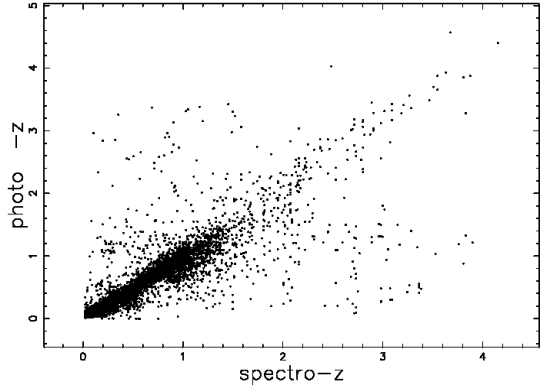}
\end{center}
\caption{
Comparison between  photo-z and  spectro-z.
}
 \label{swire_difference}%
\end{figure}

The number of galaxies of SWIRE  with 
photometric redshift  is   6095.
The differences between photometric
and spectroscopic redshift  are outlined
in Figure~\ref{swire_difference}
and a comparison should be made with Figure~10
in  \cite{Beck2017}.
A {\it first} test
parametrizes  the  differences  between
the two redshifts
as
\begin{equation}
\Delta z = (spectro-z) - (photo-z) =-0.037 \pm 0.339
\quad ,
\end{equation}
where the error is the standard deviation of the sample.
A {\it second} test  fits the
photometric-spectroscopic  relationship 
with a straight line 
\begin{equation} 
y = a +b x
\quad ,
\end{equation}
where x is the spectroscopic redshift, 
y is the photometric   redshift,
a and b two parameters to be found  with the least square fit
\cite{press}.
In our case  $a=0.102$ , $b=0.778$ and the correlation coefficient, $r$,
is 0.778.

\subsection{The observed LF}

A LF for galaxies   is  built
according to the following points:
\begin{enumerate}
\item
An   upper value of redshift
is  chosen (i.e. $0.05$);
\item
SWIRE's  galaxies are selected according
to the following  ranges of existence:
$7  \leq Log10(L/L{\sun}) \leq 11$ where $L$  is the
B-band luminosity;
\item  We organize a histogram
with bins large $\approx$ one decade;
\item
We then divide the  obtained frequencies by the
involved comoving volume;
\item
We   apply  the  $1/V_{max}$ method because
our sample is incomplete at low values of luminosity,
see \cite{Avni1980,Eales1993,Ellis1996}.
The maximum value in redshift at which a galaxy
can be detected 
is  found solving the following non linear 
equation 
\begin{equation}
f_{ave} = \frac {L_i}{4\pi \dl(z;c,H_0,\om)^2} 
\quad ,
\end{equation} 
where $f_{ave}$ is the averaged flux
of the sample and $L_i$ the luminosity of the considered bin $i$;
\item
The error  of the LF is evaluated as the square root
of the frequencies divided by the  comoving  volume,
as given by eqn.~(\ref{comovingvolume}).
\end{enumerate}
An example of LFs for SWIRE that are found by implementing  the
$1/V_{max}$
 estimator, see \cite{Schmidt1968}, can
be found in Figure 2 in \cite{Onyett2006}.

\subsection{Statistical tests}

The merit function $\chi^2$
is  computed as
\begin{equation}
\chi^2 =
\sum_{j=1}^n ( \frac {LF_{theo} - LF_{astr} } {\sigma_{LF_{astr}}})^2
\quad ,
\label{chisquare}
\end{equation}
where      $n$ is the number of bins for the LF of the galaxies,
the index  $theo$ stands  for `theoretical',
the index  $astr$ stands  for `astronomical'
and  $\sigma_{LF_{astr}}$ is the error in the LF.
The reduced  merit function $\chi_{red}^2$
is  evaluated  by
\begin{equation}
\chi_{red}^2 = \chi^2/NF
\quad,
\label{chisquarereduced}
\end{equation}
where $NF=n-k$ is the number of degrees  of freedom
and $k$ is the number of parameters.
The goodness  of the fit can be expressed by
the probability $Q$, see  Equation 15.2.12  in \cite{press},
which involves the number of degrees of freedom
and $\chi^2$.
According to  \cite{press}, the
fit ``may be acceptable'' if  $Q \geq 0.001$.
The Akaike information criterion
(AIC), see \cite{Akaike1974},
is defined by
\begin{equation}
AIC  = 2k - 2  ln(L)
\quad,
\end {equation}
where $L$ is
the likelihood  function  and $k$ is  the number of  free parameters
in the model.
We assume  a Gaussian distribution for  the errors
and we also assume that the likelihood  function
can be derived  from the $\chi^2$ statistic
$L \propto \exp (- \frac{\chi^2}{2} ) $
where  $\chi^2$ has been computed by
Equation~(\ref{chisquare}),
see~\cite{Liddle2004}, \cite{Godlowski2005}.
The AIC now becomes
\begin{equation}
AIC  = 2k + \chi^2
\quad.
\label{AIC}
\end {equation}

\subsection{Schechter LF}

\label{secschechter}
The Schechter LF of galaxies ,$\Phi$,
see \cite{schechter}, is

\begin{equation}
\Phi (L;\Phi^*,\alpha,L^*) dL =
(\frac{\Phi^*}{L^*}) (\frac{L}{L^*})^{\alpha}
\exp \bigl ({- \frac{L}{L^*}} \bigr ) dL \quad,
\label{lf_schechter}
\end{equation}
where $\alpha$ sets the slope for low values
of $L$,
$L^*$ is the
characteristic luminosity, and $\Phi^*$ represents
the number of galaxies  per $Mpc^3$.
The  normalization is
\begin{equation}
\int_0^{\infty} \Phi (L;\Phi^*,\alpha,L^*) dL  =
\rm \Phi^*\, \Gamma \left( \alpha+1 \right)
\quad  ,
\label{norma_schechter}
\end{equation}
where
\begin{equation}
\rm \Gamma \, (z )
=\int_{0}^{\infty}e^{{-t}}t^{{z-1}}dt
\quad ,
\end{equation}
is the Gamma function.
The average luminosity,
${\langle L \rangle } $, is
\begin{equation}
{\langle (\Phi (L;\Phi^*,\alpha,L^*) \rangle}
=
\rm L^* \,{\rm \Phi^*  }\,\Gamma \left( \alpha+2 \right)
\quad.
\label{ave_schechter}
\end{equation}
An equivalent form  in absolute magnitude
of the Schechter LF
is
\begin{eqnarray}
\Phi (M;\Phi^*,\alpha,M^*)dM=
\nonumber\\
0.921 \Phi^* 10^{0.4(\alpha +1 ) (M^*-M)}
\exp \bigl ({- 10^{0.4(M^*-M)}} \bigr)  dM \, ,
\label{lfstandard}
\end{eqnarray}
where $M^*$ is the characteristic magnitude.
We briefly recall the existence 
of the Wisconsin-Indiana-Yale-NOAO  Observatory (WIYN)
at Kitt Peak National Observatory.

\begin{figure}
\begin{center}
\includegraphics[width=7cm]{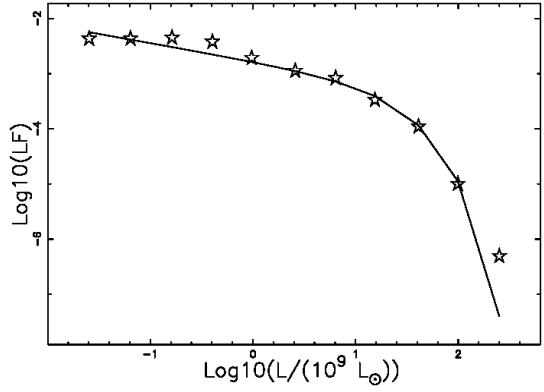}
\end{center}
\caption{
The  observed SWIRE/WIYN $LF_{24}$ for galaxies, empty stars,
(data extracted by the author) and the fit  by  the Schechter LF.
}
 \label{lf_schechter_noder_articolo}%
\end{figure}

A typical result of  the Schechter LF in the case
of  SWIRE/WIYN $LF_{24}$, see Figure2 in  \cite{Onyett2006}
where the index 24 stands for $24\mu\,m$, 
is  reported  in Figure~\ref{lf_schechter_noder_articolo}
with parameters as in Table~\ref{schechterfit}.

\subsection{The lognormal  LF}

\label{seclognorm}

Let $L$ be a random variable taking
values $L$ in the interval
$[0, \infty]$; the {\em lognormal}.
The PDF,
following \cite{evans}
or formula (14.2)$^\prime$ in
\cite{univariate1}, is
\begin{equation}
PDF (L;L^*,\sigma) = \frac
{
\sqrt {2}{{\rm e}^{-\frac{1}{2}\,{\frac {1}{{\sigma}^{2}} \left( \ln  \left( {
\frac {L}{{\it L^*}}} \right)  \right) ^{2}}}}
k}
{
2\,L\sigma\,\sqrt {\pi }
}
\quad,
\label{pdflognormal}
\end{equation}
where
$L^*=\exp{\mu_{LN}}$ and
$\mu_{LN}=\ln {L^*}$.
The mean luminosity is
\begin{equation}
E(L;L^*,\sigma)=
{\it L^*}\,{{\rm e}^{\frac{1}{2}\,{\sigma}^{2}}}
\quad .
\end{equation}

The luminosity function for galaxies, $\Phi (L;L^*,\sigma)$,
can be obtained
by  multiplying  the lognormal PDF by
$\Phi^*$, which is the number of galaxies  per Mpc$^3$ units
\begin{equation}
\Phi (L;L^*,\sigma) =\Phi^* \frac
{
\sqrt {2}{{\rm e}^{-\frac{1}{2}\,{\frac {1}{{\sigma}^{2}} \left( \ln  \left( {
\frac {L}{{\it L^*}}} \right)  \right) ^{2}}}}
}
{
2\,L\sigma\,\sqrt {\pi }
}
\quad ,
\label{pdflognormalgal}
\end{equation}
for further details, see \cite{Zaninetti2016c}.
The magnitude version for the lognormal LF
is
\begin{equation}
\Phi (M;M^*,\sigma) =
 0.3674\,{\frac {{\it \Phi^*  }}{\sigma}{{\rm e}^{- 0.4241\,
{\frac { \left( {\it M^*}- 1.0\,{\it M} \right) ^{2}}{{\sigma}^{2}
}}}}}
\quad ,
\label{pdflognormalgalmag}
\end{equation}
where $M^*$  is  the scaling absolute magnitude
and   $M$      is  the absolute magnitude.

\begin{figure}
\begin{center}
\includegraphics[width=7cm]{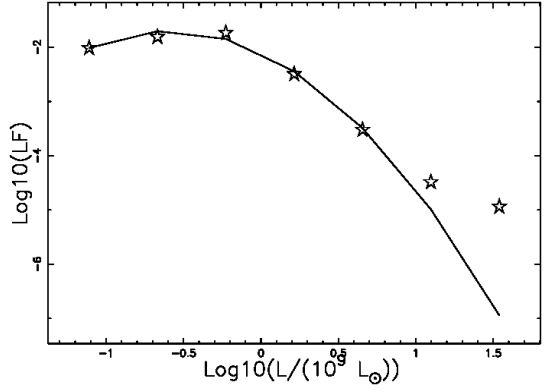}
\end{center}
\caption{
The  observed LF for galaxies in B-band , empty stars with error bar,
and the fit  by  the lognormal LF
when the distance covers the   $[0 , 0.05 ]$ range
in redshift.
}
 \label{swire_lf_lognorm_mio}%
\end{figure}

The resulting fitting  curve
is displayed in Figure  \ref{swire_lf_lognorm_mio},
with parameters as in the second column of
Table \ref{lognormfit}.

\begin{table}[ht!]
\caption
{
Parameters
and statistical tools
for   SWIRE LF
and   SWIRE/WIYN $LF_{24}$
as modeled by the
lognormal LF
for  $z$ in $[0,0.05]$ with the SWIRE  data.
}
\label{lognormfit}
\begin{center}
\begin{tabular}{|c|c|c|}
\hline
Parameter  &  SWIRE \,LF & SWIRE/WIYN $LF_{24}$ \\
\hline
$\frac{L^*}{10^{9}\,L_{\sun}}$       & 0.696  & 3.25  \\
$\sigma$                             & 0.996 & 1.77  \\
$\frac {\Phi^*}{Mpc^{-3}}$           & 0.021 & 0.014 \\
Q                                    & 0     & 2.05\,$10^{-6}$   \\
NF                                   & 4     &  4     \\
$\chi_{red}^2$                       & 59.56 & 5.12   \\
AIC                                  & 238.27& 47.03  \\
\hline
\end{tabular}
\end{center}
\end{table}

\begin{table}[ht!]
\caption
{
Parameters
and statistical tools
for   SWIRE LF
and   SWIRE/WIYN $LF_{24}$
as modeled by the
Schechter LF
for  $z$ in $[0,0.05]$ with the SWIRE  data.
}
\label{schechterfit}
\begin{center}
\begin{tabular}{|c|c|c|}
\hline
Parameter  &  SWIRE \,LF & SWIRE/WIYN $LF_{24}$ \\
\hline
$\frac{L^*}{10^{9}\,L_{\sun}}$       & 1.035   & 28.46  \\
$\alpha$                             & -0.02  & -0.33  \\
$\frac {\Phi^*}{Mpc^{-3}}$           & 0.02 & 0.0156 \\
Q                                    & 0      & $6.977\,10^{-5}$  \\
NF                                   & 4      & 4      \\
$\chi_{red}^2$                       & 126 & 4.087   \\
AIC                                  & 510 & 38.69  \\
\hline
\end{tabular}
\end{center}
\end{table}

\begin{figure}
\begin{center}
\includegraphics[width=7cm]{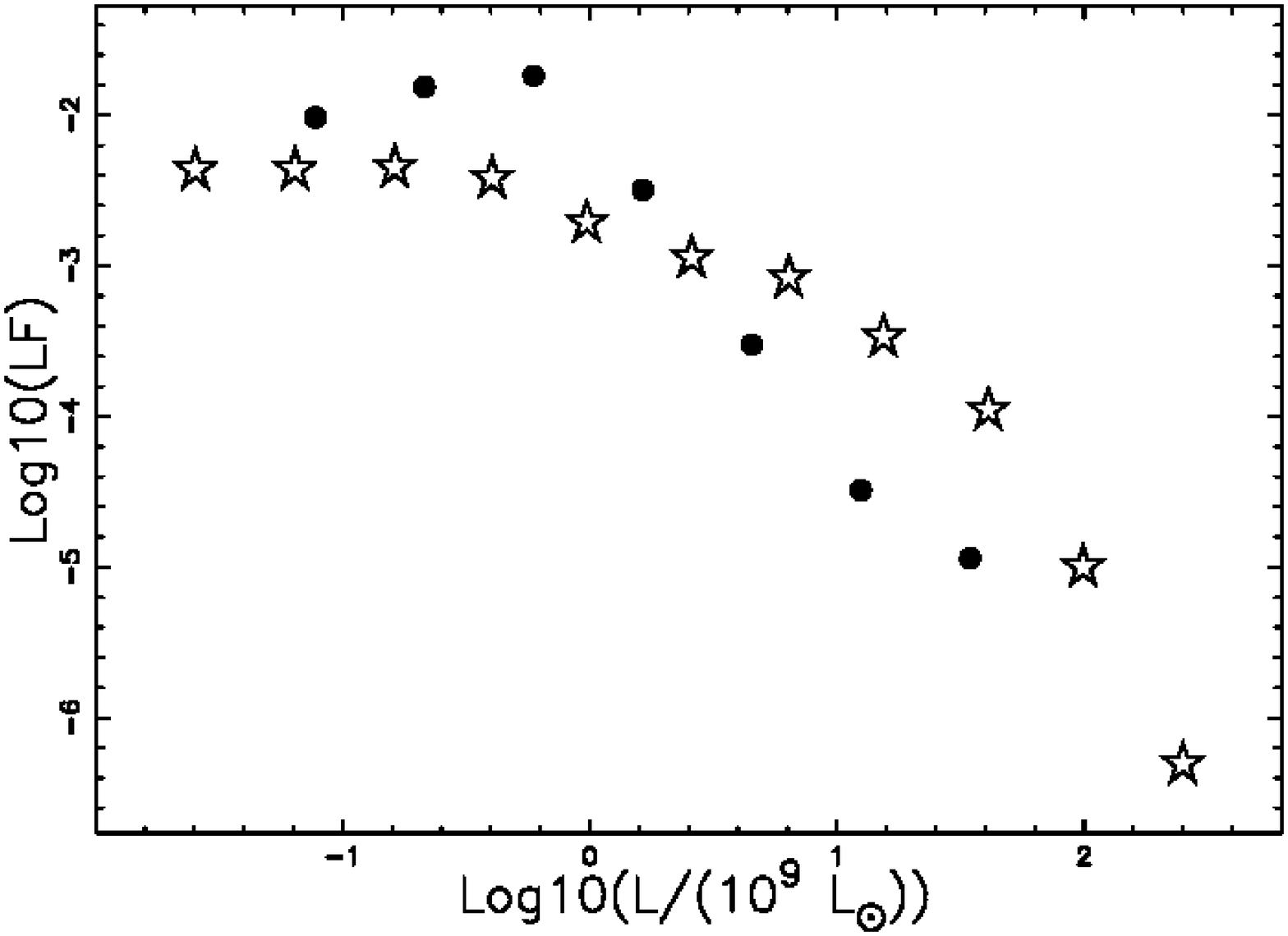}
\end{center}
\caption
{
SWIRE/WIYN $LF_{24}$,
empty stars (data extracted by the author),
and SWIRE LF in B-band, filled circles.
}
 \label{swire_lf_due}%
\end{figure}

A comparison can be made with the SWIRE/WIYN 24$\mu\,m$
LF, $LF_{24}$,
as reported in Figure 2 in  \cite{Onyett2006} (data extracted by the
author),
see Figure
\ref{swire_lf_due}.

\begin{figure}
\begin{center}
\includegraphics[width=7cm]{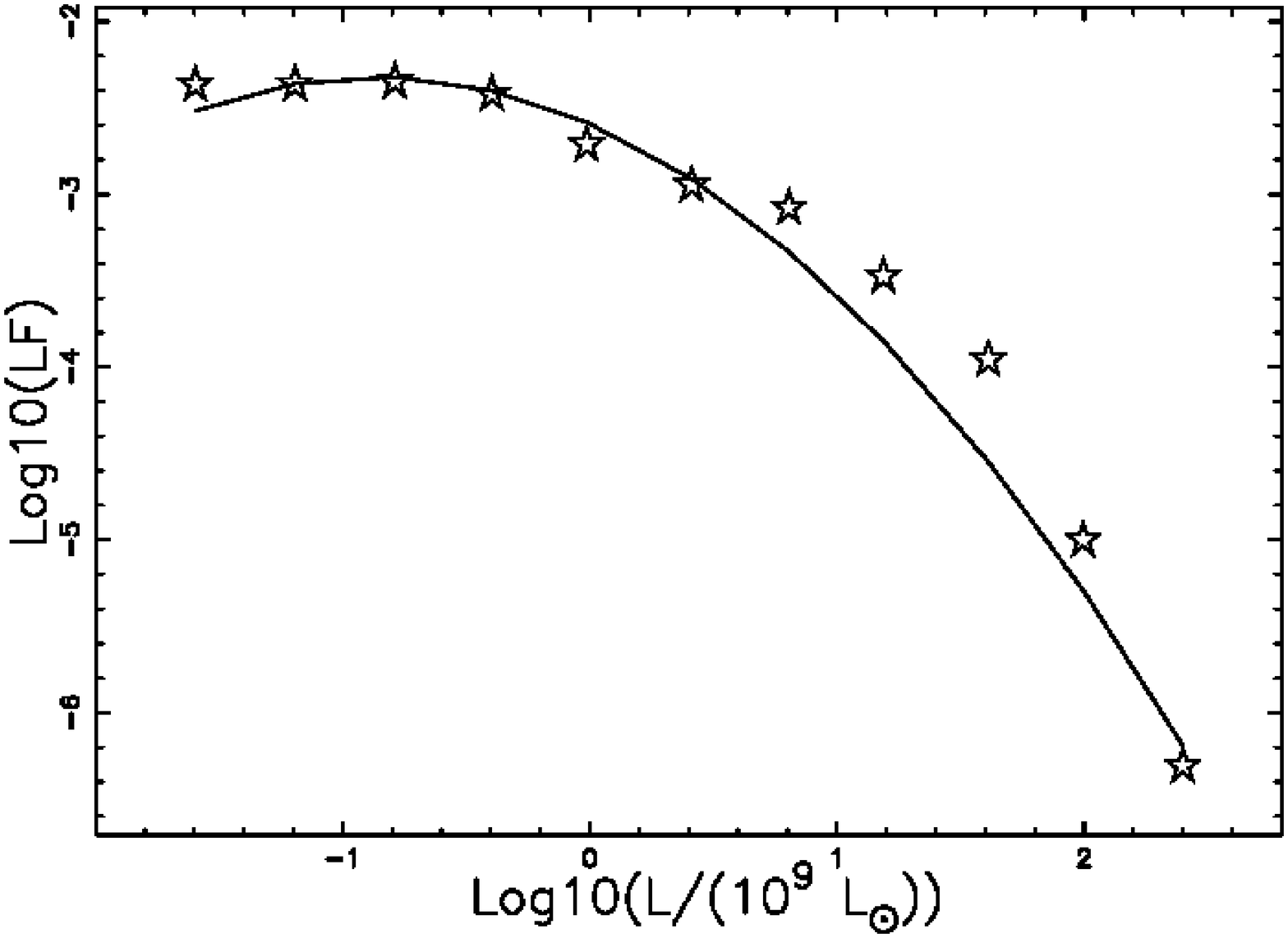}
\end{center}
\caption{
The  observed SWIRE/WIYN $LF_{24}$ for galaxies, empty stars,
(data extracted by the author)
and the fit  by  the lognormal  LF.
}
 \label{lf_lognorm_articolo}%
\end{figure}

A typical result of  the lognormal  LF in the case
of  SWIRE/WIYN $LF_{24}$
is  reported  in Figure~\ref{lf_lognorm_articolo},
with parameters as in the second column of
Table \ref{lognormfit}.

\section{Photometric maximum}

\label{sec_maximum}
The flux,$f$,
is
\begin{equation}
f  = \frac{L}{4 \pi r^2}
\quad ,
\label{flux}
\end{equation}
where $r$ is the luminosity  distance.
The redshift  is approximated as
\begin{equation}
 z =z_{2,2}
\quad ,
\end{equation}
where  $z_{2,2}$ has  been introduced
into equation~(\ref{z22}).
 The relationship between $dr$
 and $dz$ is
\begin{equation}
dr =  \frac
{
104306\, \left(  2676.6\,z+ 772.8914\,{z}^{2}+
 846.892 \right)
}
{
\left( z+ 83.4793 \right) ^{2} \left( z+ 1.7548 \right) ^{2}
}
dz
\quad  ,
\end{equation}
where $r$ has been defined as $\ml$
 by the  minimax rational approximation.
The joint distribution in {\it z}
and  {\it f}  for the number of galaxies
 is
\begin{equation}
\frac{dN}{d\Omega dz df} =
\frac{1}{4\pi}\int_0^{\infty} 4 \pi r^2 dr
\Phi (L;L^*,\sigma)
\delta\bigl(z- (z_{2,2})\bigr)
\delta\bigl(f-\frac{L}{4 \pi r^2}    \bigr)
\quad ,
\label{nfunctionzlognorm}
\end{equation}
where $\delta$ is the Dirac delta function
and  $\Phi (L;L^*,\sigma)$  has  been defined
in equation~(\ref{pdflognormalgal}).
This formula has the following explicit version
\begin{eqnarray}
\frac{dN}{d\Omega dz df} =
{ 2.92\times 10^{19}}\,{\frac { \left( z+ 0.6317 \right)
^{2} \left( z+ 0.001599 \right) ^{2} \left( z+ 3.1108
 \right)  \left( z+ 0.35222 \right) }{ \left( z+ 83.4793
 \right) ^{4} \left( z+ 1.7548 \right) ^{4}f\sigma}}
\times \nonumber  \\
{
{{\rm e}^{-
 0.5\,{\frac {1}{{\sigma}^{2}} \left( \ln  \left(
 1.141\,10^{13}\,{\frac {f \left( z+ 0.63179 \right) ^{2}
 \left( z+ 0.00159987 \right) ^{2}}{ \left( z+ 83.4793
 \right) ^{2} \left( z+ 1.75483 \right) ^{2}{\it L^*}}} \right)
 \right) ^{2}}}}}
\quad .
\label{nzflat}
\end{eqnarray}

\begin{figure}
\begin{center}
\includegraphics[width=6cm]{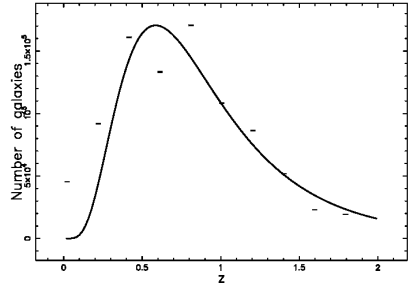}
\end{center}
\caption
{
The galaxies of SWIRE in B-band  with
$ 0.00057
 \, L_{\sun}/Mpc^2  \leq
f \leq  10501 \, L_{\sun}/Mpc^2  $
are  organized by frequency versus
distance (empty circles);
the error bar is given by the square root of the frequency.
The maximum frequency of the observed galaxies is at
 $z=0.9 $.
The full line is the theoretical curve
generated by
$\frac{dN}{d\Omega dr df}$
as given by the application of the lognormal  LF,
which  is Equation~(\ref{nzflat}),
and the theoretical  maximum is at
$z=0.57 $.
The parameters are
$L^*= 9 \,10^{10} L_{\sun}$   and
$\sigma$ =1.4 \,.
}
          \label{swire_maximum}%
    \end{figure}

\begin{figure}
\begin{center}
\includegraphics[width=10cm]{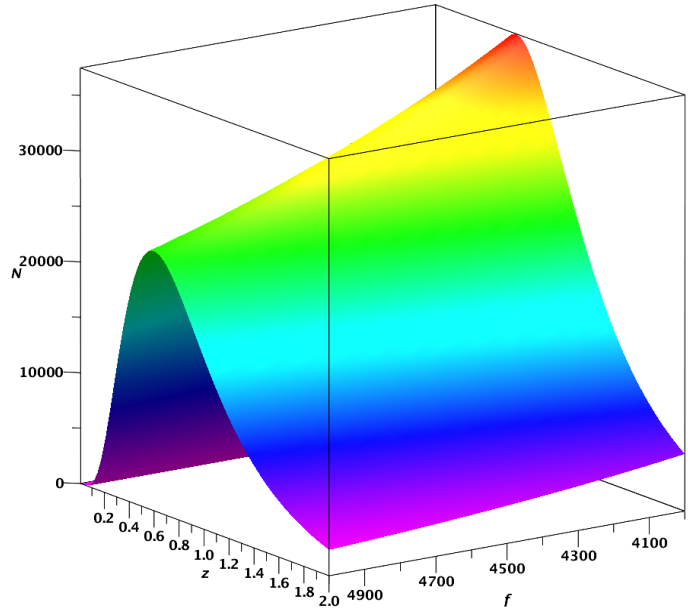}
\end{center}
\caption
{
The theoretical number of galaxies
as a function of redshift and  flux expressed
in  $L_{\sun}/Mpc^2$.
The parameters are
$\frac{L^*}{10^{9}\,L_{\sun}}=9\,10^9$,
$\sigma=1.4$
and
$\frac {\Phi^*}{Mpc^{-3}}=0.015$ .
}
          \label{fluxz}%
    \end{figure}

\begin{figure}
\begin{center}
\includegraphics[width=10cm]{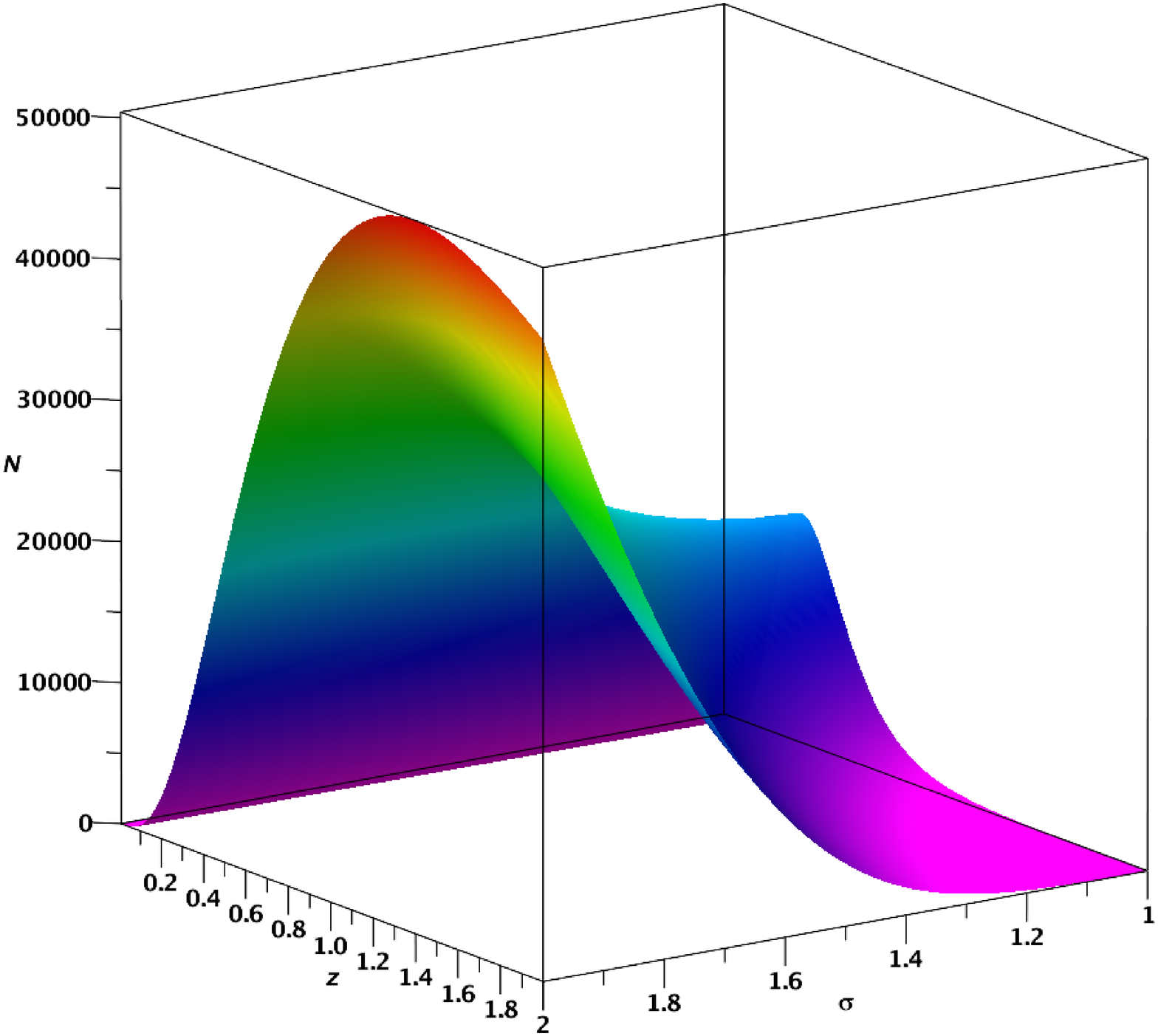}
\end{center}
\caption
{
The theoretical number of galaxies
as a function of $sigma$ and redshift
when
$\frac{L^*}{10^{9}\,L_{\sun}}=9\,10^9$,
$f=5250 L_{\sun}/Mpc^2 $
and
$\frac {\Phi^*}{Mpc^{-3}}=0.015$ .
}
          \label{sigmaz}%
    \end{figure}

Figure~\ref{swire_maximum}
presents the number of  galaxies  that are observed in  SWIRE
as a function  of the redshift  for  a given
window in  flux, in addition to the theoretical curve.
The theoretical number of galaxies
is reported in Figure~\ref{fluxz}
as a function of the flux  and redshift,
and is reported in Figure~\ref{sigmaz}
as a function of $\sigma$ and redshift.

An analogous procedure for the
joint distribution in {\it z}
and  {\it f}  for the number of galaxies
in the case of the
Schechter   LF
derives
\begin{eqnarray}
\frac{dN}{d\Omega dz df} =
{ 8.3533\times 10^{32}}\,{\frac { \left( z+ 0.001599
 \right) ^{4} \left( z+ 0.6317 \right) ^{4}{\it \Phi^*}\,
 \left( z+ 3.110 \right)  \left( z+ 0.3522 \right) }{
 \left( z+ 83.4793 \right) ^{6} \left( z+ 1.75483 \right) ^{6}
{\it L^*}}
}\times
\nonumber  \\
{
\left(  1.141\,10^{13}\,{\frac {f \left( z+
 0.63179 \right) ^{2} \left( z+ 0.001599 \right) ^{2}}{
 \left( z+ 83.4793 \right) ^{2} \left( z+ 1.7548 \right) ^{2}
{\it L^*}}} \right) ^{\alpha}{{\rm e}^{- 1.141 \,10^{13}\,{\frac {f
 \left( z+ 0.631791 \right) ^{2} \left( z+ 0.001599 \right)
^{2}}{ \left( z+ 83.4793 \right) ^{2} \left( z+ 1.7548
 \right) ^{2}{\it L^*}}}}}}
\label{nzflatschechter}
\end{eqnarray}

\begin{figure}
\begin{center}
\includegraphics[width=6cm]{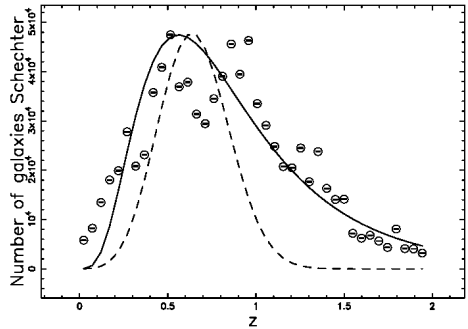}
\end{center}
\caption
{
The galaxies of SWIRE in B-band
in
frequencies  versus
redshift  (empty circles).
The full line is the theoretical curve
generated by
$\frac{dN}{d\Omega dr df}$,
as given by the application of the lognormal  LF,
which  is Equation~(\ref{nzflat}),
with parameters as  in Figure~\ref{swire_maximum}.
The dashed line is the theoretical curve
generated by
$\frac{dN}{d\Omega dr df}$,
as given by the application of the Schechter  LF,
which  is given in Equation~(\ref{nzflatschechter}),
with parameters as given in Figure~\ref{swire_maximum}.
The input parameters of the Schechter LF
are
$\frac{L^*}{10^{9}\,L_{\sun}}$=200,
$\alpha$=-0.33 and
$\frac {\Phi^*}{Mpc^{-3}}$=0.0156.
The
goodness of fit is
$\chi^2= 389246$  for the Schechter LF    and
$\chi^2= 83970$   for the lognormal LF.
}
          \label{swire_maximum_two}%
    \end{figure}

\section{Conclusions}

{\bf Flat cosmology}
In this paper, we derive an approximate relationship
for the luminosity distance
in spatially flat cosmology with pressure-less matter and
cosmological constant, $H_0 = 70  \h0units$ and $\om=0.277$,
as a function of the redshift, see equation (\ref{dlpade22}).
We have derived the inverse relationship,
the redshift as a function of the luminosity distance,
see equation (\ref{z22}).

{\bf spectro-z vs. photo-z}
The differences between  spectroscopic and photometric redshift,
as  processed from the SWIRE catalog,
are
characterized  by a nearly  zero value,
-0.037,  and  by a great error,  0.339.
This  fact demands for an improvement in the derivation
of the photometric redshift \cite{Soo2018}.

{\bf LF}
Knowledge of the comoving volume allows us
to derive the astronomical LF for  galaxies of
the SWIRE catalog in the framework of the photometric redshift.
The LF is modeled with lognormal LF,
see Section \ref{seclognorm},
and a comparison is made with the
Schechter LF, see Section \ref{secschechter}.
We have  analyzed  two observed LFs
with two theoretical LFs
for a total of four cases.
The $\chi_{red}^2$ of the Schechter LF
was  smaller  than that of the  lognormal
in one case over four, see Tables
\ref{lognormfit}  and  \ref{schechterfit}.
In the case of the PM, the lognormal  number of
galaxies as a function of the redshift
gives a better fit in respect to the  Schechter fit,
see $\chi^2$ in Figure \ref{swire_maximum_two}.

{\bf PM}

The joint distribution in the photometric redshift and in the
energy flux density
is modeled in the case of a flat  universe
and a lognormal LF,
see Formula (\ref{nfunctionzlognorm}).
The position in the redshift of the maximum for PM
of galaxies
at  a given flux or apparent magnitude
does not have an analytical expression
and, therefore, is found numerically,
see Figure \ref{swire_maximum}.

\section*{Acknowledgments}

This research has made use of the VizieR catalogue access tool, CDS,
Strasbourg, France.


\begin{thebibliography}{10}
\expandafter\ifx\csname url\endcsname\relax
  \def\url#1{\texttt{#1}}\fi
\expandafter\ifx\csname urlprefix\endcsname\relax\def\urlprefix{URL }\fi
\expandafter\ifx\csname href\endcsname\relax
  \def\href#1#2{#2} \def\path#1{#1}\fi

\bibitem{Rowan-Robinson2013}
M.~{Rowan-Robinson}, E.~{Gonzalez-Solares}, M.~{Vaccari}, L.~{Marchetti},
  {Revised SWIRE photometric redshifts}, \mnras 428 (2013) 1958--1967.
\newblock \href {http://arxiv.org/abs/1210.3471} {\path{arXiv:1210.3471}},
  \href {http://dx.doi.org/10.1093/mnras/sts163}
  {\path{doi:10.1093/mnras/sts163}}.

\bibitem{Dalya2016}
G.~{Dalya}, Z.~{Frei}, G.~{Galgoczi}, P.~{Raffai}, R.~S. {de Souza}, {VizieR
  Online Data Catalog: GLADE catalog (Dalya+, 2016)}, VizieR Online Data
  Catalog 7275.

\bibitem{Karasev1982}
B.~V. {Karasev}, {Statistical genesis of a lognormal distribution as a source
  of properties observed in the clumping of galaxies}, Pisma v Astronomicheskii
  Zhurnal 8 (1982) 527--534.

\bibitem{Mcbreen1994}
B.~{McBreen}, K.~J. {Hurley}, R.~{Long}, L.~{Metcalfe}, {Lognormal
  Distributions in Gamma-Ray Bursts and Cosmic Lightning}, \mnras 271 (1994)
  662.
\newblock \href {http://dx.doi.org/10.1093/mnras/271.3.662}
  {\path{doi:10.1093/mnras/271.3.662}}.

\bibitem{Li1996}
H.~{Li}, E.~E. {Fenimore}, {Log-normal Distributions in Gamma-Ray Burst Time
  Histories}, \apjl 469 (1996) L115.
\newblock \href {http://arxiv.org/abs/astro-ph/9607131}
  {\path{arXiv:astro-ph/9607131}}, \href {http://dx.doi.org/10.1086/310275}
  {\path{doi:10.1086/310275}}.

\bibitem{Nakar2001}
E.~{Nakar}, T.~{Piran}, {New Results on the Temporal Structure of GRBs}, in:
  E.~{Costa}, F.~{Frontera}, J.~{Hjorth} (Eds.), Gamma-ray Bursts in the
  Afterglow Era, 2001, p. 348.
\newblock \href {http://arxiv.org/abs/astro-ph/0103011}
  {\path{arXiv:astro-ph/0103011}}.

\bibitem{Ioka2002}
K.~{Ioka}, T.~{Nakamura}, {A Possible Origin of Lognormal Distributions in
  Gamma-Ray Bursts}, \apjl 570 (2002) L21--L24.
\newblock \href {http://arxiv.org/abs/astro-ph/0202053}
  {\path{arXiv:astro-ph/0202053}}, \href {http://dx.doi.org/10.1086/340815}
  {\path{doi:10.1086/340815}}.

\bibitem{Zaninetti2016c}
L.~{Zaninetti}, {The Truncated Lognormal Distribution as a Luminosity Function
  for SWIFT-BAT Gamma-Ray Bursts}, Galaxies 4 (2016) 57.
\newblock \href {http://arxiv.org/abs/1611.01650} {\path{arXiv:1611.01650}},
  \href {http://dx.doi.org/10.3390/galaxies4040057}
  {\path{doi:10.3390/galaxies4040057}}.

\bibitem{Hurley1995}
K.~J. {Hurley}, B.~{McBreen}, M.~{Delaney}, A.~{Britton}, {Lognormal Properties
  of SGR 1806-20 and Implications for Other SGR Sources}, \apss 231 (1995)
  81--84.
\newblock \href {http://arxiv.org/abs/astro-ph/9508074}
  {\path{arXiv:astro-ph/9508074}}, \href {http://dx.doi.org/10.1007/BF00658592}
  {\path{doi:10.1007/BF00658592}}.

\bibitem{McBreen1998}
B.~{McBreen}, K.~J. {Hurley}, {Lognormal properties of SGR1806-20 and the
  possibility of a quiescent population of other SGR sources}, in: C.~A.
  {Meegan}, R.~D. {Preece}, T.~M. {Koshut} (Eds.), Gamma-Ray Bursts, 4th
  Hunstville Symposium, Vol. 428 of American Institute of Physics Conference
  Series, 1998, pp. 939--943.
\newblock \href {http://arxiv.org/abs/astro-ph/9807218}
  {\path{arXiv:astro-ph/9807218}}, \href {http://dx.doi.org/10.1063/1.55418}
  {\path{doi:10.1063/1.55418}}.

\bibitem{Marr2015a}
J.~H. {Marr}, {Angular momentum of disc galaxies with a lognormal density
  distribution}, \mnras 453 (2015) 2214--2219.
\newblock \href {http://arxiv.org/abs/1507.04515} {\path{arXiv:1507.04515}},
  \href {http://dx.doi.org/10.1093/mnras/stv1734}
  {\path{doi:10.1093/mnras/stv1734}}.

\bibitem{Marr2015b}
J.~H. {Marr}, {Galaxy rotation curves with lognormal density distribution},
  \mnras 448 (2015) 3229--3241.
\newblock \href {http://arxiv.org/abs/1502.02949} {\path{arXiv:1502.02949}},
  \href {http://dx.doi.org/10.1093/mnras/stv216}
  {\path{doi:10.1093/mnras/stv216}}.

\bibitem{schechter}
P.~{Schechter}, {An analytic expression for the luminosity function for
  galaxies.}, \apj 203 (1976) 297--306.

\bibitem{Zaninetti2008}
L.~{Zaninetti}, {A new luminosity function for galaxies as given by the
  mass-luminosity relationship }, \aj 135 (2008) 1264--1275.

\bibitem{Zaninetti2010f}
L.~{Zaninetti}, {The Luminosity Function of Galaxies as modelled by the
  Generalized Gamma Distribution }, Acta Physica Polonica B 41~(4) (2010)
  729--751.

\bibitem{Bilicki2016}
M.~{Bilicki}, J.~A. {Peacock}, T.~H. {Jarrett}, et~al., {WISE * SuperCOSMOS
  Photometric Redshift Catalog: 20 Million Galaxies over 3/pi Steradians},
  \apjs 225 (2016) 5.
\newblock \href {http://arxiv.org/abs/1607.01182} {\path{arXiv:1607.01182}},
  \href {http://dx.doi.org/10.3847/0067-0049/225/1/5}
  {\path{doi:10.3847/0067-0049/225/1/5}}.

\bibitem{Varela2012}
J.~{Varela}, J.~{Betancort-Rijo}, I.~{Trujillo}, E.~{Ricciardelli}, {The
  Orientation of Disk Galaxies around Large Cosmic Voids}, \apj 744 (2012) 82.
\newblock \href {http://arxiv.org/abs/1109.2056} {\path{arXiv:1109.2056}},
  \href {http://dx.doi.org/10.1088/0004-637X/744/2/82}
  {\path{doi:10.1088/0004-637X/744/2/82}}.

\bibitem{Etherington1933}
I.~M.~H. {Etherington}, {On the Definition of Distance in General Relativity.},
  Philosophical Magazine 15.

\bibitem{Beck2017}
R.~{Beck}, C.-A. {Lin}, E.~E.~O. {Ishida}, et~al., {On the realistic validation
  of photometric redshifts}, \mnras 468 (2017) 4323--4339.
\newblock \href {http://dx.doi.org/10.1093/mnras/stx687}
  {\path{doi:10.1093/mnras/stx687}}.

\bibitem{press}
W.~H. {Press}, S.~A. {Teukolsky}, W.~T. {Vetterling}, B.~P. {Flannery},
  {Numerical Recipes in FORTRAN. The Art of Scientific Computing}, Cambridge
  University Press, Cambridge, UK, 1992.

\bibitem{Avni1980}
Y.~{Avni}, J.~N. {Bahcall}, {On the simultaneous analysis of several complete
  samples - The V/Vmax and Ve/Va variables, with applications to quasars}, \apj
  235 (1980) 694--716.
\newblock \href {http://dx.doi.org/10.1086/157673} {\path{doi:10.1086/157673}}.

\bibitem{Eales1993}
S.~{Eales}, {Direct construction of the galaxy luminosity function as a
  function of redshift}, \apj 404 (1993) 51--62.
\newblock \href {http://dx.doi.org/10.1086/172257} {\path{doi:10.1086/172257}}.

\bibitem{Ellis1996}
R.~S. {Ellis}, M.~{Colless}, T.~{Broadhurst}, J.~{Heyl}, K.~{Glazebrook},
  {Autofib Redshift Survey - I. Evolution of the galaxy luminosity function},
  \mnras 280 (1996) 235--251.
\newblock \href {http://arxiv.org/abs/astro-ph/9512057}
  {\path{arXiv:astro-ph/9512057}}, \href
  {http://dx.doi.org/10.1093/mnras/280.1.235}
  {\path{doi:10.1093/mnras/280.1.235}}.

\bibitem{Schmidt1968}
M.~{Schmidt}, {Space Distribution and Luminosity Functions of Quasi-Stellar
  Radio Sources}, \apj 151 (1968) 393.
\newblock \href {http://dx.doi.org/10.1086/149446} {\path{doi:10.1086/149446}}.

\bibitem{Onyett2006}
N.~{Onyett}, S.~{Oliver}, G.~{Morrison}, F.~{Owen}, F.~{Pozzi}, D.~{Carson},
  {SWIRE Team}, {The 24{$\mu$}m Luminosity Function of spectroscopic SWIRE
  sources from the Lockman Validation Field}, in: L.~{Armus}, W.~T. {Reach}
  (Eds.), Astronomical Society of the Pacific Conference Series, Vol. 357 of
  Astronomical Society of the Pacific Conference Series, Astronomical Society
  of the Pacific, 2006, p. 275.
\newblock \href {http://arxiv.org/abs/astro-ph/0503444}
  {\path{arXiv:astro-ph/0503444}}.

\bibitem{Akaike1974}
H.~{Akaike}, {A new look at the statistical model identification}, IEEE
  Transactions on Automatic Control 19 (1974) 716--723.

\bibitem{Liddle2004}
A.~R. {Liddle}, {How many cosmological parameters?}, \mnras 351 (2004)
  L49--L53.

\bibitem{Godlowski2005}
W.~{Godlowski}, M.~{Szydowski}, {Constraints on Dark Energy Models from
  Supernovae}, in: M.~{Turatto}, S.~{Benetti}, L.~{Zampieri}, W.~{Shea} (Eds.),
  1604-2004: Supernovae as Cosmological Lighthouses, Vol. 342 of Astronomical
  Society of the Pacific Conference Series, Astronomical Society of the
  Pacific, 2005, pp. 508--516.

\bibitem{evans}
M.~{Evans}, N.~{Hastings}, B.~{Peacock}, Statistical Distributions - third
  edition, John Wiley \& Sons Inc, New York, 2000.

\bibitem{univariate1}
N.~L. {Johnson}, S.~{Kotz}, N.~{Balakrishnan}, {Continuous univariate
  distributions. Vol. 1. 2nd ed.}, {Wiley }, New York, 1994.

\bibitem{Soo2018}
J.~Y.~H. {Soo}, B.~{Moraes}, B.~{Joachimi}, W.~{Hartley}, O.~{Lahav},
  A.~{Charbonnier}, M.~{Makler}, M.~E.~S. {Pereira}, J.~{Comparat}, T.~{Erben},
  A.~{Leauthaud}, H.~{Shan}, L.~{Van Waerbeke}, {Morpho-z: improving
  photometric redshifts with galaxy morphology}, \mnras 475 (2018) 3613--3632.
\newblock \href {http://arxiv.org/abs/1707.03169} {\path{arXiv:1707.03169}},
  \href {http://dx.doi.org/10.1093/mnras/stx3201}
  {\path{doi:10.1093/mnras/stx3201}}.

\end{thebibliography}

\end{document}